\documentclass[aps,prb,twocolumn,float,amsmath]{revtex4}
\usepackage[dvips]{color}
\usepackage[normalem]{ulem} 
 \usepackage{hyperref}
\definecolor{g-blue}{rgb}{0.83,0.95,1}
\definecolor{g-yellow}{rgb}{1,1,0.7}
\definecolor{g-green}{rgb}{0.9,1,0.9}
\definecolor{green}{rgb}{0,0.6,0}
\definecolor{cyan}{rgb}{0,0.7,0.7}
\definecolor{black}{rgb}{0,0,0}
\definecolor{grey}{rgb}{0.4 ,0.4 ,0.4 }

\usepackage{bm}
\usepackage{amssymb}
\usepackage{amsfonts}
\usepackage{amsmath}
 \usepackage{graphicx}
  \usepackage{amsmath,bm,epsfig}
\def \ed {\end{document}}
\def\Fbox#1{\vskip1ex\hbox to 8.5cm{\hfil\fboxsep0.3cm\fbox{%
  \parbox{8.0cm}{#1}}\hfil}\vskip1ex\noindent}  

\newcommand{\Eq}[1]{Eq.\,(\ref{#1})}
\newcommand{\Eqs}[1]{Eqs.\,(\ref{#1})}
\newcommand{\Fig}[1]{Fig.\,\ref{#1}}

\def\be{\begin{equation}}\def\ee{\end{equation}}
\def\bea{\begin{eqnarray}}\def\eea{\end{eqnarray}}
\def\bse{\begin{subequations}}\def\ese{\end{subequations}}
\newcommand{\BE}[1]{\begin{equation}\label{#1}}
\newcommand{\BEA}[1]{\begin{eqnarray}\label{#1}}
\newcommand{\BSE}[1]{\begin{subequations}\label{#1}}

\let \= \equiv \let\*\cdot \let\~\widetilde \let\^\widehat \let\-\overline

  \def\1{\bm1} 

\def\<{\left\langle}    \def\>{\right\rangle}
\def\({\left(}          \def\){\right)}
 \def \[ {\left [} \def \] {\right ]}

\newcommand{\B}[1]{{\bm{#1}}}
\newcommand{\C}[1]{{\mathcal{#1}}}    
\newcommand{\BC}[1]{\bm{\mathcal{#1}}}

\renewcommand{\sb}[1]{_{\text {#1}}}  
\renewcommand{\sp}[1]{^{\text {#1}}}  
\def\Sb#1{_{\scriptscriptstyle\rm{#1}}}

\def\He4 {$^4$He~}

\begin{document}
\title{Dynamics of the Density of Quantized  Vortex-Lines in Superfluid Turbulence}

 \author{D. Khomenko$^1$, L. Kondaurova$^2$,  V.S. L'vov$^1$,  P. Mishra$^1$, A.  Pomyalov$^1$ and I.  Procaccia$^1$}

 \address{$^1$Department of Chemical Physics, The Weizmann Institute of Science, Rehovot 76100, Israel \\
 $^2$Institute of Thermophysics, Novosibirsk, Russia}
\begin{abstract}
The quantization of vortex lines in superfluids requires the introduction of their density $\C L(\B r,t)$ in the description of quantum turbulence. The  space homogeneous balance equation for $\C L(t)$, proposed by Vinen on the basis of dimensional and physical considerations, allows a number of competing forms for the production term $\C P$. Attempts to choose the correct one on the basis of time-dependent homogeneous experiments ended inconclusively. To overcome this difficulty we announce here an approach that employs an inhomogeneous channel flow which is excellently suitable to distinguish the implications of the various possible forms of the desired equation. We demonstrate  that the originally selected form which was extensively used in the literature is
 in strong contradiction with our data. We therefore present a new inhomogeneous equation for $\C L(\B r,t)$ that is in agreement with our data and propose that it should be considered for further studies of superfluid turbulence.
\end{abstract}

\maketitle

 \textbf{\emph{Background}}: Below the Bose-Einstein condensation temperature $T_\lambda\approx 2.18\,$K, liquid \He4 becomes a quantum inviscid superfluid\cite{b:a,b:b,b:Annett,b:SS-2012}. Aside from the lack of viscosity,
 the vorticity in \He4 is constrained to vortex-line singularities of fixed circulation $\kappa= h/M$, where $h$ is Planck's constant  and $M$ is the mass of the \He4 atom. These vortex lines have
 a core radius $a_0\approx  10^{-8}\,$cm, compatible with the interatomic distance. In generic turbulent states, these vortex lines appear as a complex tangle with a typical intervortex distance  $\ell$\,\cite{nem}.

  {Recent progress in laboratory \cite{Helsinki-rev,b:PNAC2,b:PNAC1,b:PNAC3,b:PNAC4,b:11,B:12,B:13, B:14, B:15, B:19, B:20, B:21,B:22,B:23,b:17, b:18} and numerical studies\,\cite{b:SS-2012,b:PNAC2,b:PNAC5,b:PNAC6,B:9,b:Kond-1,b:PNAC7,b:A-10} of superfluid turbulence in superfluid $^3$He and $^4$He led to} a growing consensus  that the statistical properties of superfluid turbulence at large scales $R\gg \ell$ are similar to those of classical turbulence. An  acceptable theory of these large scale properties\,\cite{LNV-2004,LNS-2006,LNR-2008,BLPP-2012,BLPP-2013,b:PNAC2, b:PNAC1,BLNNPP-2015} is based on the Landau-Tisza two-fluid model\,\cite{b:L-41,b:T-40}, using `normal' and `superfluid' components of densities $\rho\sb n$ and $\rho\sb s$ with velocity fields $\B u\sb n(\B r,t)$ and $\B u\sb s(\B r,t)$.  {This model was extended by Hall-Vinen\cite{HV} and Bekarevich-Khalatnikov\cite{BK} to include a mutual friction between the components, proportional to the  the vortex-line density $\C L$.
 This means that a theory  of large-scale motions which is affected by the mutual friction requires the inclusion of the dynamics of $\C L$. }

 The situation  changes  drastically  upon considering the statistical properties on smaller scales, where the quantization of  vortex lines becomes crucial. Several statistical characteristics of the vortex tangle become essential for a consistent description. {Besides $\C L$, these characteristics} involve  the  mean-square  curvature $(\tilde S)^2$,  the vortex tangle anisotropy parameters $I_\parallel$ and $I_\ell$ introduced by Schwarz\cite{b:1}. The most important of these is the vortex line density $\C L(\B r,t)$. {\em It is expected that coupling one or more of these quantities to the variables of the two fluid   {Hall-Vinen-Bekarevich-Khalatnikov equations\,\cite{HV,BK}} is a minimal requirement for an acceptable theory of quantum turbulence.}

 \begin{figure*}
 \begin{tabular}{ccc}
  \includegraphics[scale=0.27]{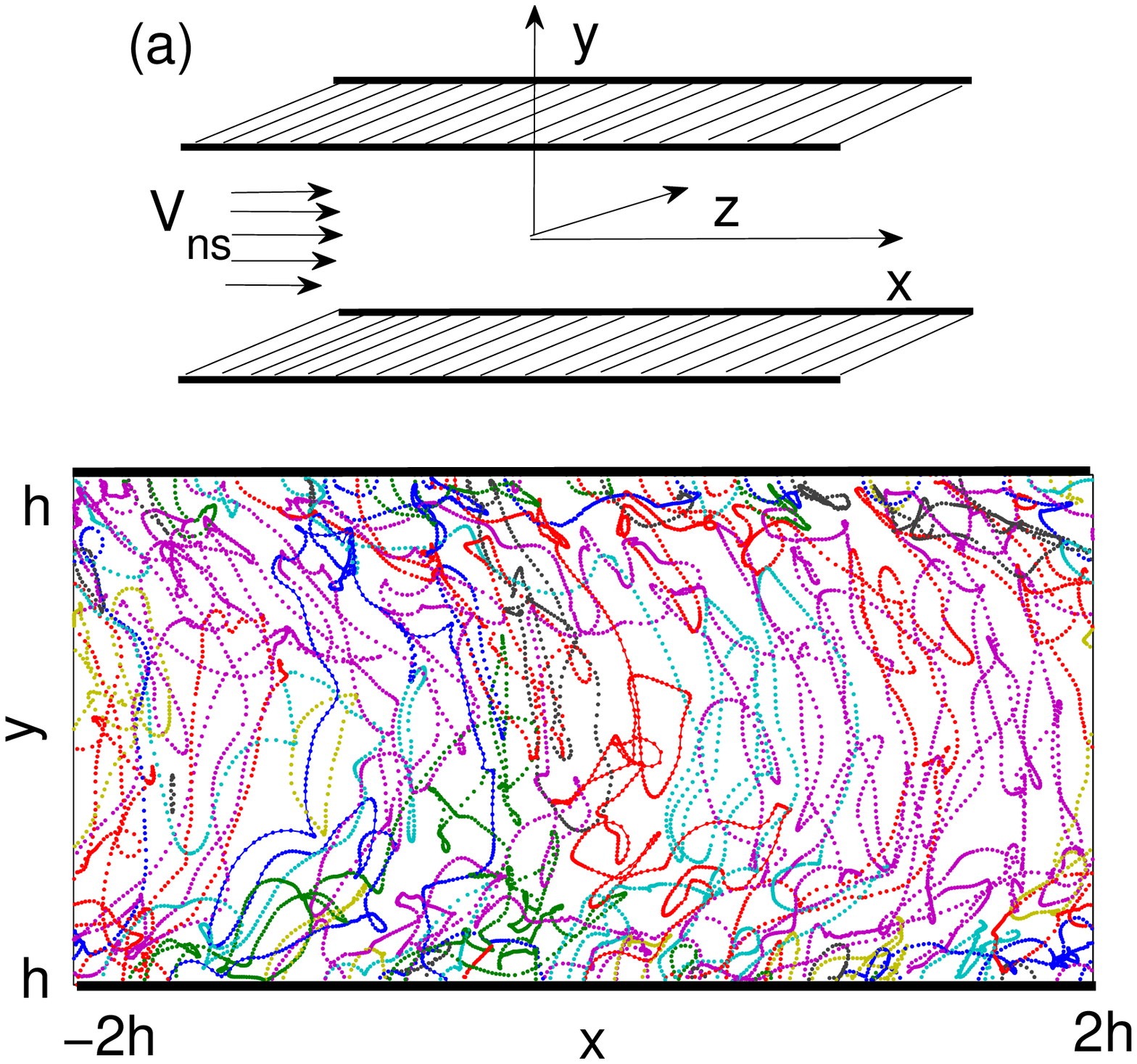}& \includegraphics[scale=0.31 ]{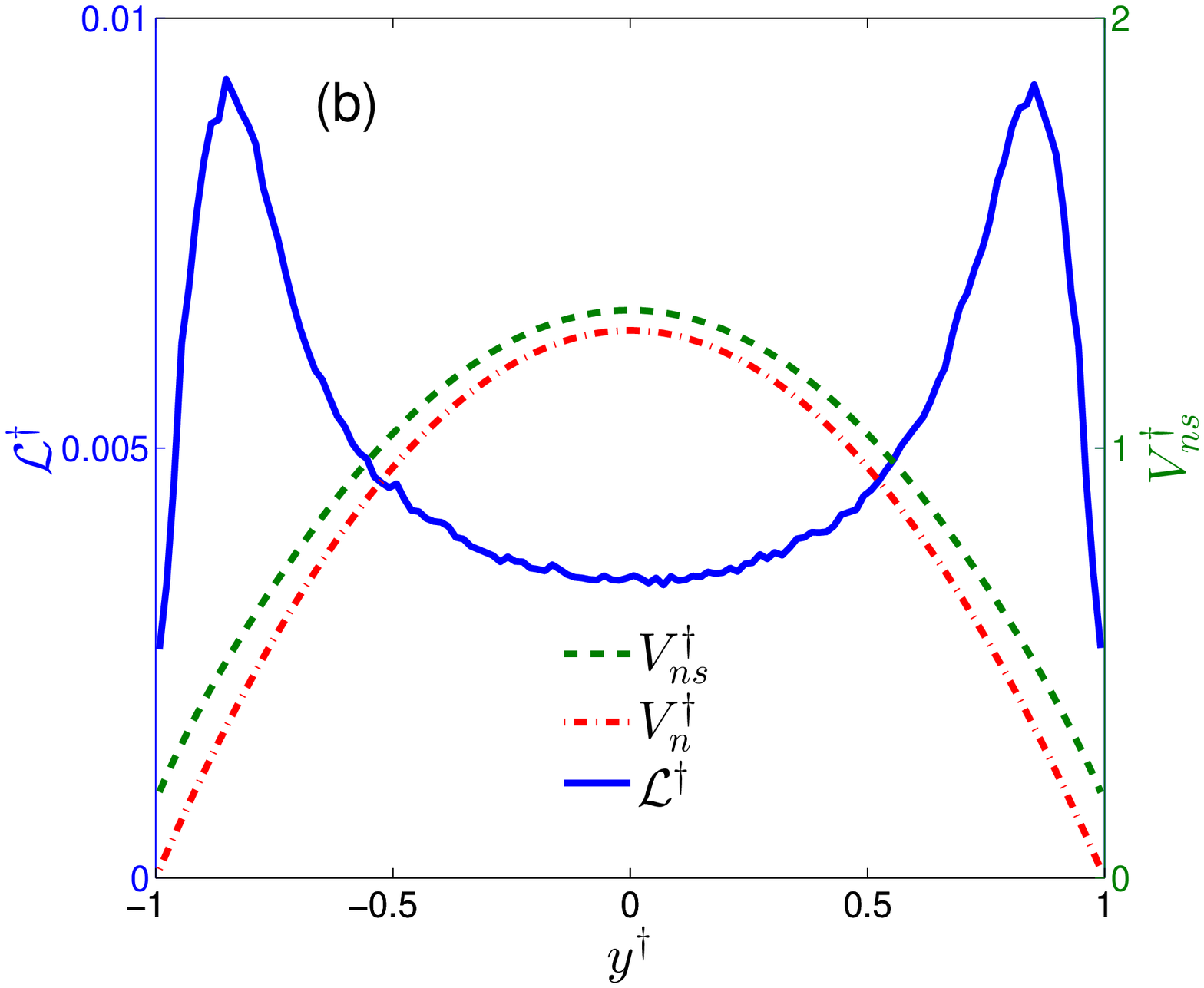} & \includegraphics[scale=0.31 ]{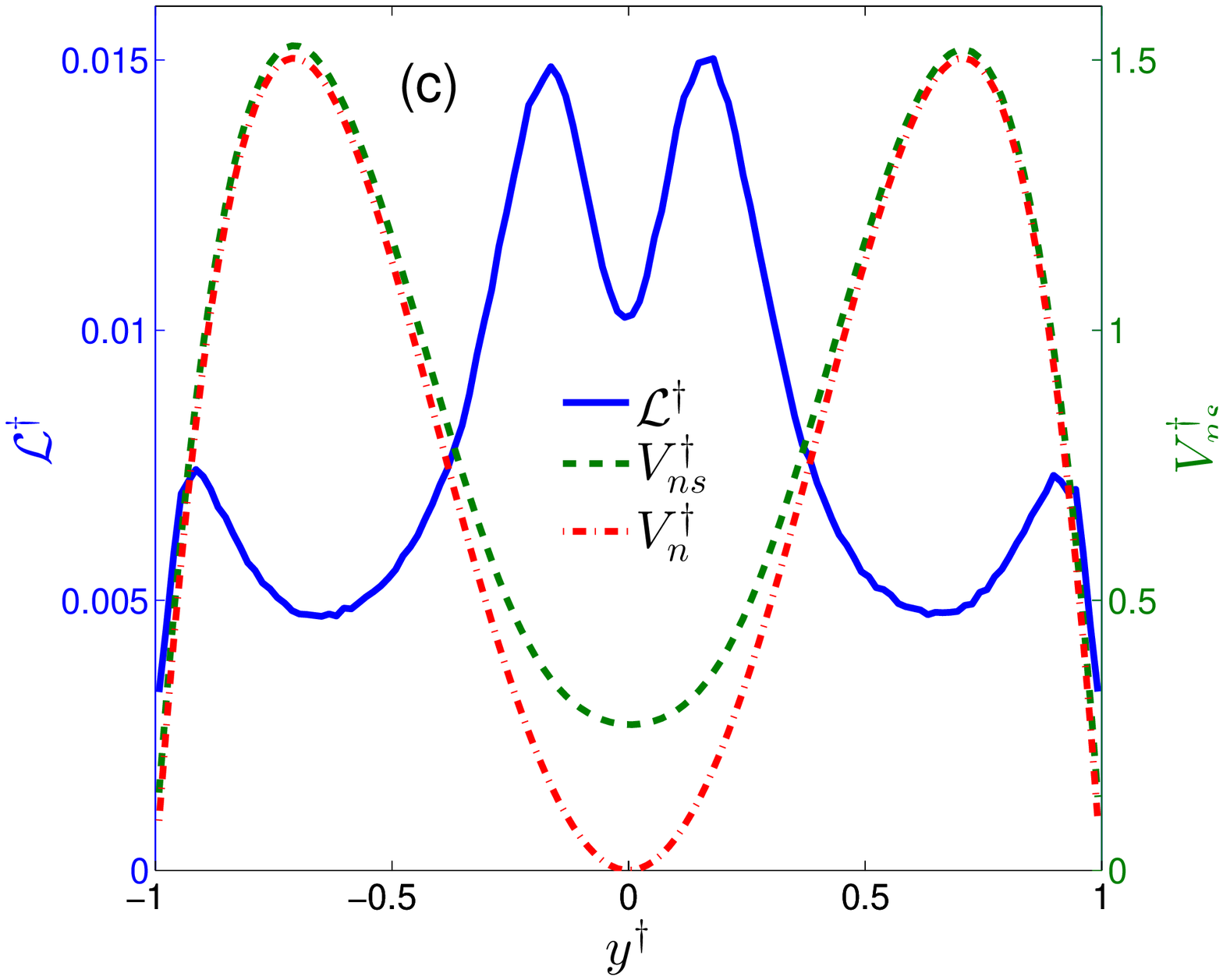} \\
  \end{tabular}
 \caption{  \label{f:1} Numerical setup and obtained profiles of counterflow velocity and vortex line density. Panel a: Upper figure: Plane channel flow geometry  $L_x\times L_y\times L_z$, $L_x=4h,L_y=2h,L_z=2h, h=0.05$~cm. Lower figure: Projection of a  vortex tangle on the $(x,y)$-plane, orthogonal to the walls. Different colors are used to distinguish vortex lines. Panels b and c: Prescribed parabolic (b) and non-parabolic (c) normal velocity profiles $V \sb{n}^\dag(y^\dag)$ ($- \cdot -$), the resulting  counterflow profiles $V \sb{ns}^\dag(y^\dag)$  ($- -$)  and    the resulting profiles of $\C L^\dag(y^\dag)$  (---)   in  dimensionless unites defined by \Eqs{non-d}. $T=1.6\;$K.}
\end{figure*}

 \textbf{\emph{The problem}}: A phenomenological equation of motion for $\C L(t)$  was suggested by Vinen for homogeneous counterflows\cite{b:5,b:6} already in 1957:
\begin{subequations}\label{eq:gen-bal}
\begin{equation}\label{eq:gen-balA}
 {d \C L(t)}/{d t}= \C P(t)- \C D (t) \ .
  \end{equation}
  Here the production term $\C P(t)$ describes the growth of $\C L$ due to the extension of the vortex rings by mutual friction which is caused by the difference between the  velocities of the normal and super components (``the counterflow velocity" $V\sb{ns}$). The decay term, $\C D(t)$ is again caused by the mutual friction due to the  moving normal fluid components and is assumed to be independent of $V\sb{ns}$. Therefore both terms should be proportional to the  dimensionless  dissipative mutual friction parameter  $\alpha$. In principle it is not guaranteed that the equation for $\C L$ can be closed via $\C L$ and $V\sb {ns}$. Such a  closure   for $\C P $ and $\C D  $\cite{b:5} assumes that (beside $\alpha$,   $\kappa$ and  $V\sb{ns}$) the only relevant variable in the problem is the instantaneous value  $\C L(t)$, while  $\~ S $,   $I_\parallel $,  $I_\ell $, etc.  are to some extent unimportant.  Upon accepting this closure idea the dimensional reasoning dictates\cite{b:5,b:6,nem}:
 \begin{eqnarray}\label{eq:gen-balB}
 \C P \Rightarrow \C P\sb {cl}&=&  \alpha \, \kappa \C L^2   F(x)\,, \quad x\=V\sb{ns}^2 \big /  \kappa^2  \C L\,,  \\ \label{eq:gen-balC}
 \C D \Rightarrow \C D\sb {cl}  &=& \alpha \, \kappa \C L^2  G(x)\ .
 \end{eqnarray}
 \end{subequations}
   Here $F (x)$ and $G(x)$ are dimensionless  functions of the dimensionless argument $x$.

 The most delicate issue in this approach is the determination of the functions $F(x)$ and $G(x)$. Vinen\cite{b:5}  assumed that the decay term $\C D$ is independent of $\B V\sb{ns}$ leading to $G(x)=C_{\rm dec}$. On the other hand Vinen  {and later authors (see,  e.g. \cite{b:Kop})} chose $\C P $ to be proportional to the mutual friction force $f\propto \alpha |V\sb{ns}|$, leading to the proposition that $F(x)\propto \sqrt x$, and then
 \begin{subequations}\label{eq:Vin} \begin{equation}\label{eq:VinA}
 \C P  \sb {cl} \Rightarrow \C P_1  = \alpha  \,  C_1 \C L^{3/2} |V\sb{ns}|\,,  
 \end{equation}
  where $C_1$ is a dimensionless constant.  Vinen\cite{b:6} realized that \eqref{eq:VinA} is not the only possibility. Another choice can follow the spirit of Landau's theory of critical phenomena, considering $\C L$ as an order parameter which determines $d\C L/dt$ via an analytical function $F(x)$. Then the  leading term in the expansion of $F(x)\propto x$ giving:
  \begin{equation}\label{eq:VinB}
 \C P \sb {cl} \Rightarrow \C P_2   = \alpha \,  C_2 \C L \, V\sb{ns}^2\big / \kappa \ .
 \end{equation}
Both options\,\eqref{eq:Vin} are of course dimensionally correct and they predict the same stationary solution, $\C L\sb{st}\propto V\sb{ns}^2$, which is well supported by both experiments and numerical simulations (see, e.g. \cite{b:Kond-1} and references therein).
 \end{subequations}
 In principle, one could hope to distinguish between the different forms of this important equation by comparing their prediction for the {\em time evolution} from some initial condition toward $\C L\sb{st}$ in the presence of counterflow $V\sb {ns}$. Unfortunately, the difference in prediction is too small. Neither Vinen \cite{b:6} himself  nor later\,\cite{b:8} experimental attempts succeeded to distinguish between these two forms\,\footnote{For more detailed discussion of this problem see, e.g. \cite{nem}}. We have made our own attempts to distinguish between the two discussed models \eqref{eq:Vin} by numerical simulation of space homogeneous counterflow turbulence, finding  inconclusive results as well.

 \textbf{\emph{The proposed resolution}}  of this old conundrum can be obtained by studying {\em inhomogeneous flows} like channel flows in which the various relevant variables have nontrivial profiles. We will argue that in fact none of the equations \eqref{eq:Vin} are correct. We propose yet a third form of $\C P $ [corresponding to $f(x)\propto x^{3/2}$]:
  \begin{subequations}\label{P3}
  \begin{equation}\label{P3A}
 \C P \sb {cl}  \Rightarrow \C P_3   = \alpha \,  C\sb{prod} \sqrt{\C L} \, V\sb{ns}^3\big / \kappa^2\ .
  \end{equation}
  Being  dimensionally correct this closure {\em fits the data that are presented below
  significantly better than either of the equations \eqref{eq:Vin}}. We are led to a revision of the homogeneous equation of motion for  the field $\C L(\B r, t)$ in the form
  \begin{equation}\label{P3B}
  \frac{\partial  \C L(\B r, t)}{\partial t} + \B \nabla \cdot \BC J\sb{cl}(\B r, t)= \C P_3(\B r, t)- \C D\sb{cl} (\B r, t) \ .
   \end{equation}
   Here we have added   a  vortex-line density  flux  $\BC J (\B r, t)$.  {Based on our numerical simulations (see below)   we suggest to model $\BC J (\B r, t)$ as follows}:
 \begin{equation}\label{flux}
  \BC J\sb{cl}(\B r,t)= -C\sb{flux}\big(\alpha\big / 2 \kappa\big ) \B \nabla     V \sb {ns}  ^2   \ .
   \end{equation}
    \end{subequations}

  {Notice that the suggested \Eqs{P3} are based on our analysis of counterflow turbulence with laminar normal fluid components. Nevertheless we believe that   \em \Eqs{P3} or their modifications  may  serve as a   basis for future studies of  inhomogeneous superfluid turbulence in a wide variety of conditions,  including  the evolution of a neutron-initiated micro big bang in superfluid $^3${\rm He} \cite{BB}, turbulent counterflows and pressure-driven superfluid channel and pipe flows of $^4${\rm He}, turbulent flow of $^3${\rm He} in rotating cryostat \cite{Helsinki-rev}, etc.}.

 \begin{figure*}
 \begin{tabular}{ccc}
 \includegraphics[scale=0.32]{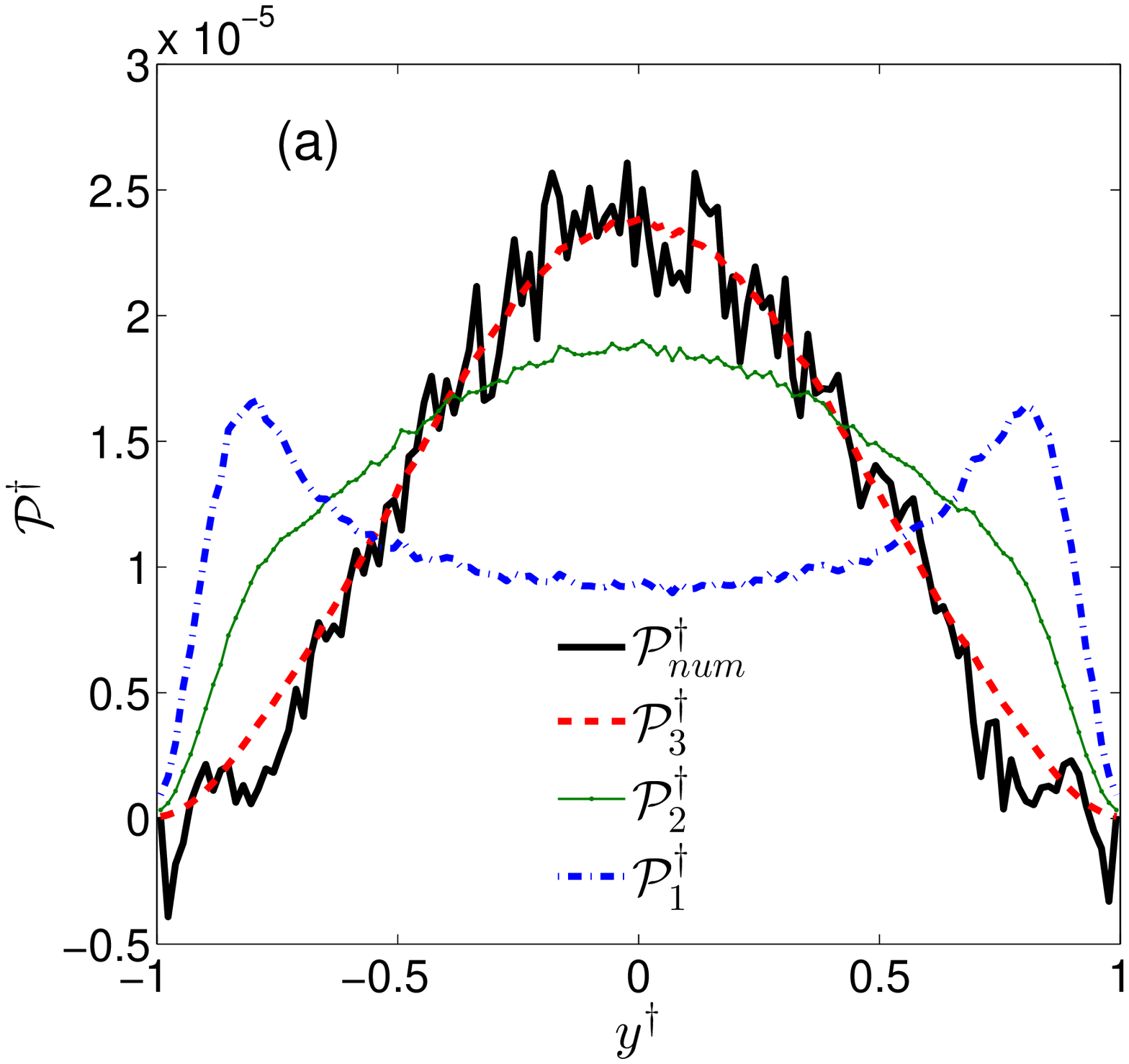} &  \includegraphics[scale=0.32]{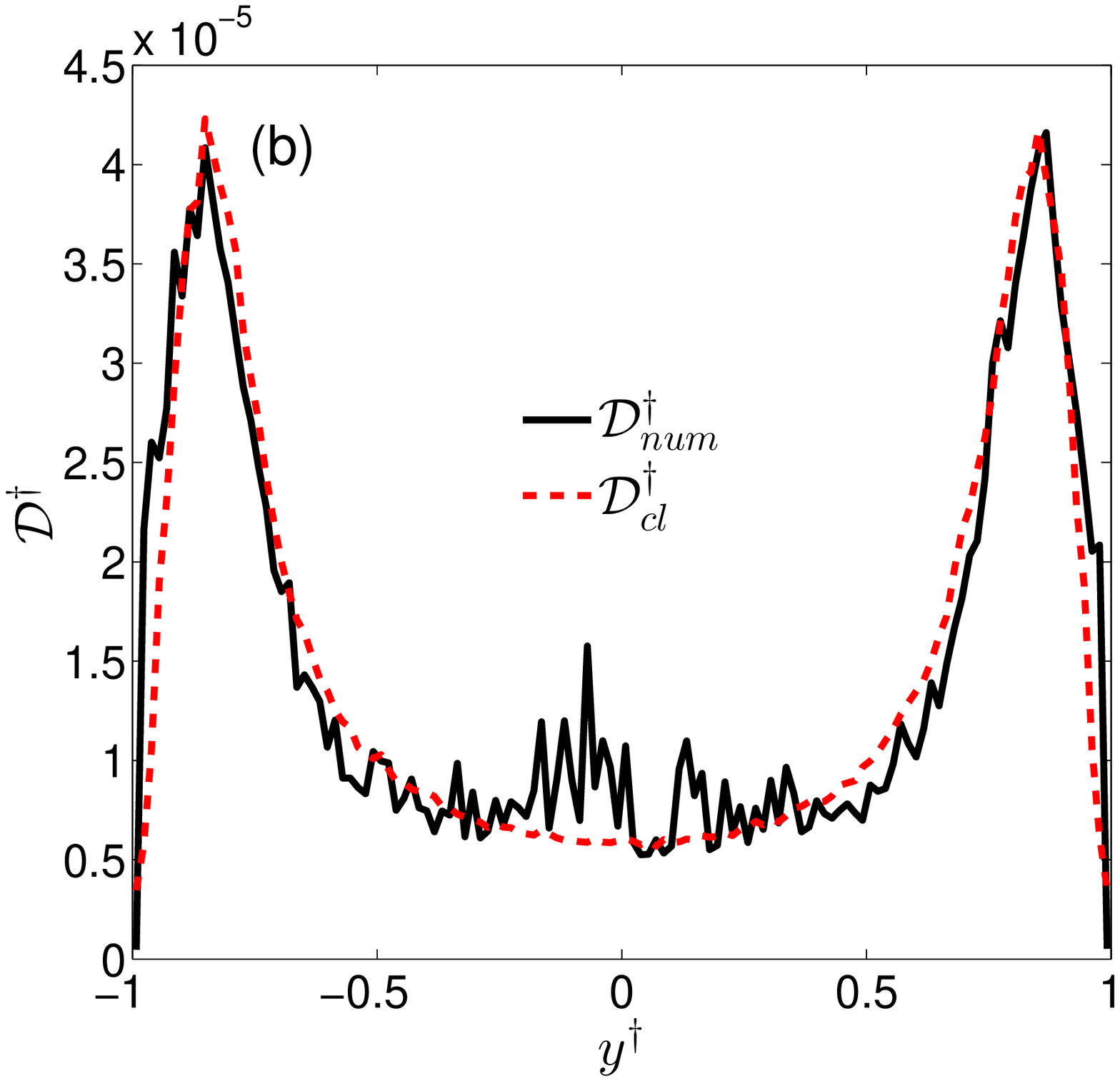} &   \includegraphics[scale=0.32]{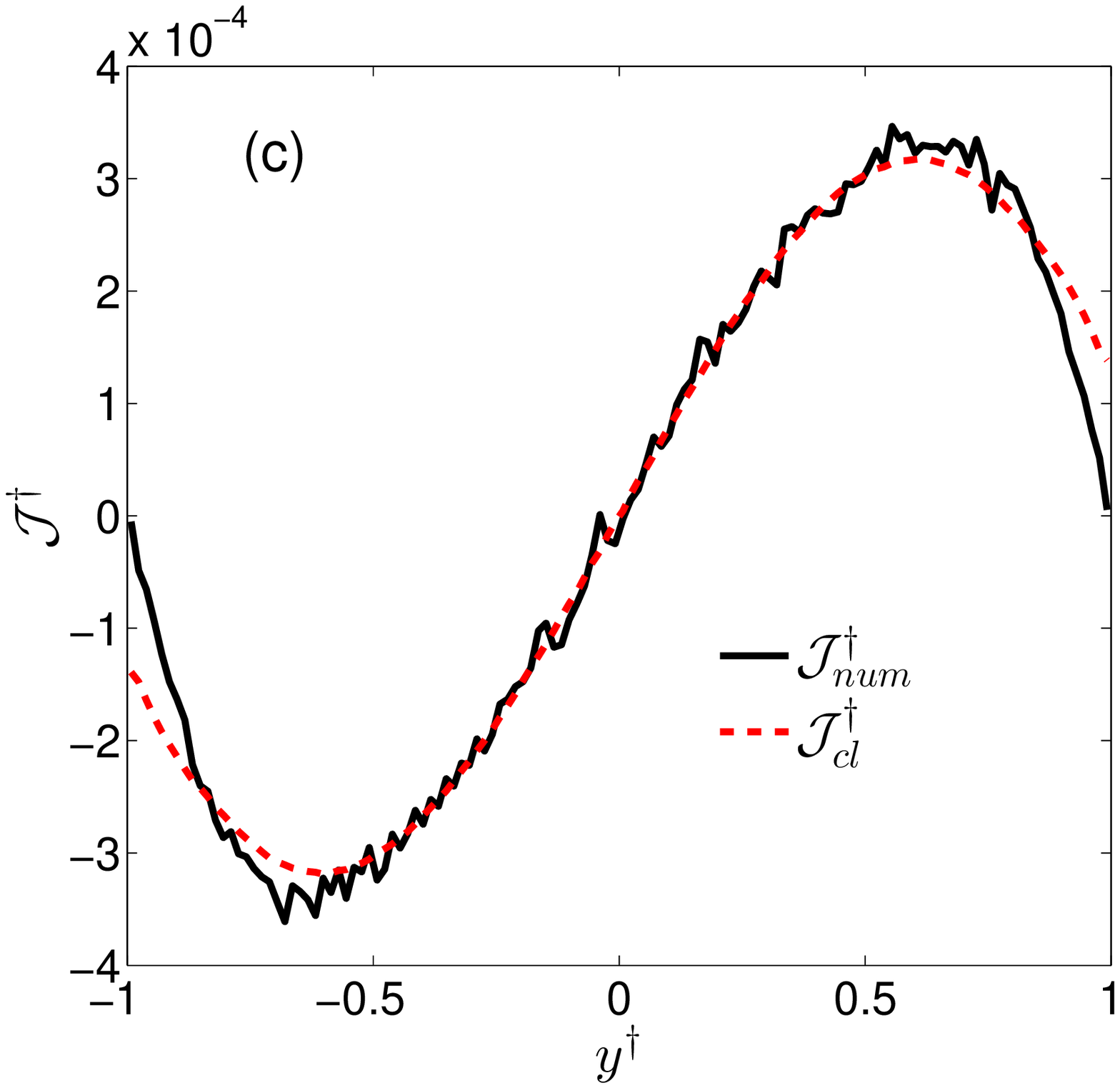}
 \end{tabular}\vskip -.3 cm
\caption{\label{f:2}
For parabolic normal velocity profile: A comparison between production ($\C P$), decay ($\C D$) and flux ($\C J$) as obtained from numerical simulation and those obtained using their closure versions. Comparison of  the numerical data ( {\Large \bf -}\!{\Large \bf -}) to the competing production forms   of the Vinen equation (Panel a): ($- \cdot -$) --$\C P_1$, \Eq{eq:VinA}, (---) -- $\C P_2$, \Eq{eq:VinB} and ($-~-$) -- $\C P_3$, \Eq{P3B}.  Comparison   of  the numerical data for decay (Panel b) and flux term (panel c) with \Eq{eq:gen-balC} for $\C D$ and \Eq{flux} for $ \C J$ --  ($- ~ -$).}

\end{figure*}

 \textbf{\emph{ Starting from first principles}}:
 To reach the desired equation (\ref{P3B}) we denote the  coordinates of the  quantized vortex lines by   $\B s(\xi,t) $, which is parameterized by the arc-length $\xi$. Schwarz\,\cite{b:1} derived the equation of motion  for the length of the vortex-line segment $\delta \xi$:
 \begin{subequations} \label{theor}\begin{equation}
  \frac{1}{\delta \xi} \frac{d\delta \xi }{dt}\approx  \alpha \B V\sb{ns}(\B s,t)\cdot(\bm s' \times \bm s'') \ . \label{dxdt2}
 \end{equation} %
  Here $ \bm  s'  =d \bm  s/d\xi $, $ \bm  s''  =d^2 \bm  s/d\xi^2 $, $\alpha$ is the temperature dependent dissipative  mutual friction parameter. The counterflow velocity $\B V\sb{ns}$ (which is a function of $\B s$ and $t$ which we suppress for notational simplicity)  is the difference between the normal fluid velocity $\B V\sp n$  and super-fluid   velocity $\B V\sp s$:
  \begin{equation}
   \B V\sb{ns} \=   {\bm V}^{\rm n}-\bm {V}^{\rm s} \,, \   \B V\sp s= \B V\sp s _0  + \B V\Sb{BS}\ .
  \end{equation}
The super-fluid velocity   $ \B V\sp s$ includes the macroscopic potential part $\B V^{\rm s}_0$, and  the Biot-Savart velocity $\B V\Sb{BS}$.  The later term is defined by the entire vortex tangle configuration $\C C$:
  \begin{eqnarray}\label{BSE}
{\B V}\Sb{BS} =   \frac{\kappa}{4\pi}\int _{ \C C} \frac{(\bm s- \bm s_1)\times \bm ds_1}{|\bm s- \bm s_1|^3} = \B V\sp s\Sb{LIA}+ \B V \sb{nl} \sp s\ .
\end{eqnarray}
The integral \eqref{BSE} is logarithmically divergent when $\B s_1 \to \B s$. It is customary to regularize it by using the vortex core radius $a_0$ and the mean vortex line curvature radius $R\= 1/\widetilde S$. The main contribution to $\B V\Sb{BS}$ originates from integrating over scales between $a_0$ and $R$,  i.e. $a_0\leqslant |\B s_1 - \B s| \leqslant R$. This contribution is known as the ``Local Induction Approximation" (LIA)\cite{b:1} and is written as:
 \begin{equation}\label{eq2A}
  \B V\sp s\Sb{LIA} =\beta \bm s' \times \bm s''\,,   \quad
  \beta\equiv (\kappa / 4\pi )\ln\big(R/a_0\big)\ .
  \end{equation}
  The $\B V\sp s\sb{nl}$ term  is non-local,    being produced by the rest of the vortex configuration, $\C C'$,  with $|\B s_1 - \B s| > R$:
 \begin{eqnarray}\label{eq2a}
   \B V\sp s\sb{nl} &=&\frac{\kappa}{4\pi}\int _{\C C'}\frac{(\bm s- \bm s_1)\times \bm ds_1}{|\bm s- \bm s_1|^3}\,\ \ .
  \end{eqnarray}   \end{subequations}

The next step is to  integrate  \Eq{dxdt2} over the vortex tangle in a fixed volume $\Omega$ which resides in slices between $y$ and $y+\delta y$, going over all $x$ and $z$.  This provides us with
 the equation of motion for $ \C L(y,t) \equiv {\int_{\C C_{\Omega}}  d\xi}/{\Omega}$.
 This equation is written in the form similar to \Eq{P3B}:
 \begin{equation}
        \frac{\partial \C L(y,t)}{\partial t} + \frac {\partial \C J (y,t)}{\partial y} =  \C P (y,t)- \C D  (y,t) \,,
\end{equation}
with the following identification for the flux $ \C J$ (toward the walls), production $ \C P $ and decay term $ \C D$:
\begin{subequations}\label{9}\begin{eqnarray}
\C P (y,t) &=& \frac{\alpha}{\Omega} \int _{\C C_{\Omega}} d\xi \,  (\B V^{\rm n} -\B V^{\rm s}_0-V_{\rm nl}^{\rm s}) \cdot (\B s'\times\B s'')\,, ~~~~~\label{prod}\\
 \C D (y,t) &=& \frac{\alpha\beta}{\Omega} \int _{\C C_{\Omega}}  d\xi \,  |\B s''|^2 = \alpha \beta \C L \widetilde S^2\,, \label{dec}\\ \label{9a}
 \C J (y,t) &=&\frac{1}{\Omega} \int _{\C C_{\Omega}} d\xi V_{{\rm drift}, y} = \frac{\alpha}{\Omega} \int _{\C C_{\Omega}} d\xi \, V\sb{ns,x} s'_z\ .
 \end{eqnarray}\end{subequations}
 Here the production and decay terms come directly from integrating  \Eq{dxdt2}; they  coincide with the corresponding equations in Ref.\cite{b:1} with the only difference that in \Eqs{9} the integrals are taken in the slice $\Omega$ (between $y$ and $y+\delta y$), while in   \cite{b:1} the integrals are taken over the entire volume. In the flux term the drift velocity $\B V\sb{drift}(\xi)$ of the vortex line segment $\B s(\xi), $,  can be found from the vortex filament equations\cite{b:1} and written as follows:
 \begin{equation}\label{drift}
 \B V\sb{drift}(\xi)= \B V\sp s+\alpha \B s' \times \B V\sb{ns}\ .
\end{equation}
 The mean value of $\B V\sp s$ is oriented along the $x$ direction and it does not contribute to the $y$ component of the flux, $\C J$. The $y$-component of the second term in \Eq{drift} gives the final expression in \Eq{9a}.

    \begin{figure*}
    \begin{tabular}{ccc}
  \includegraphics[scale=0.31 ]{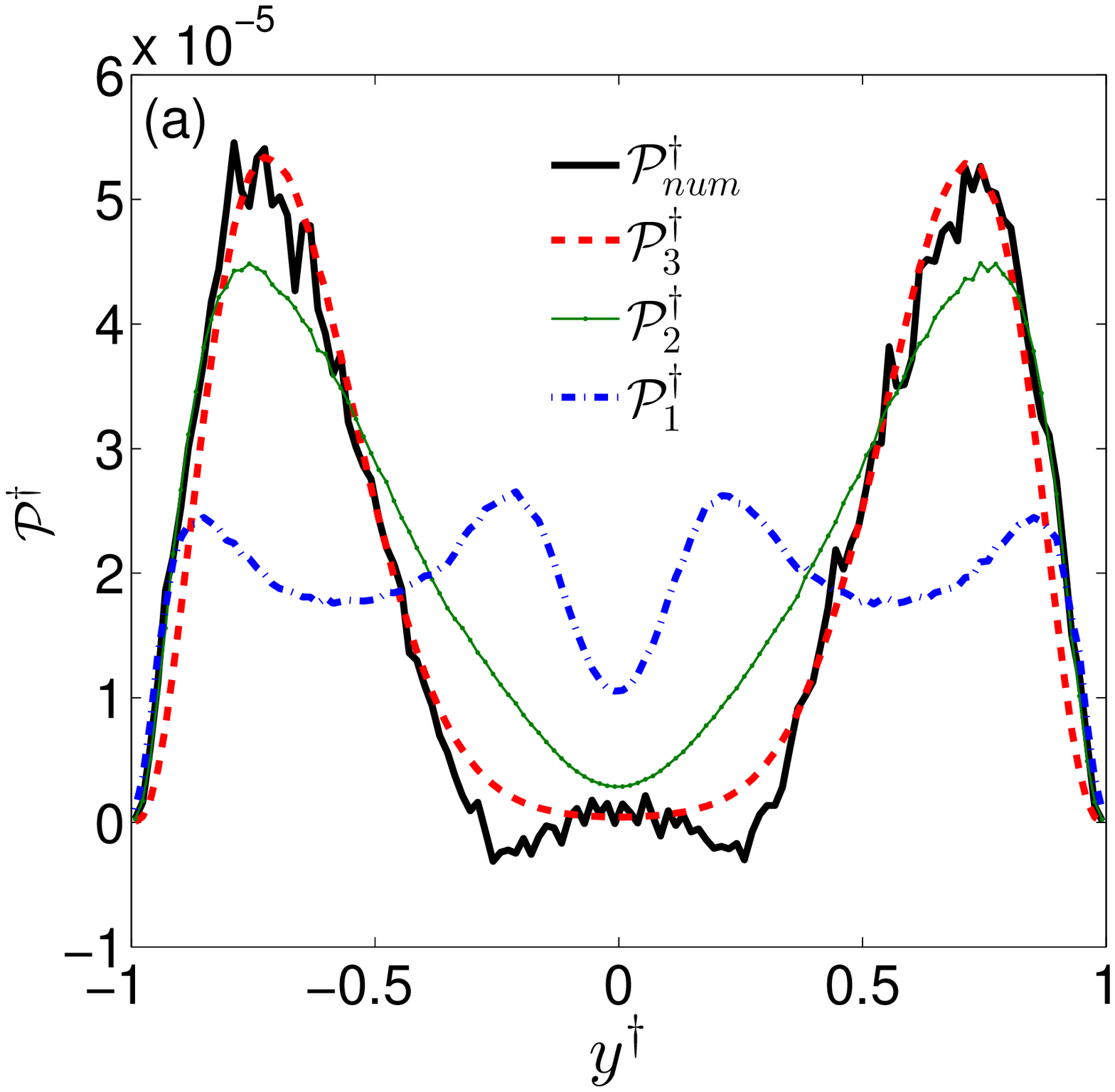} &  \includegraphics[scale=0.31 ]{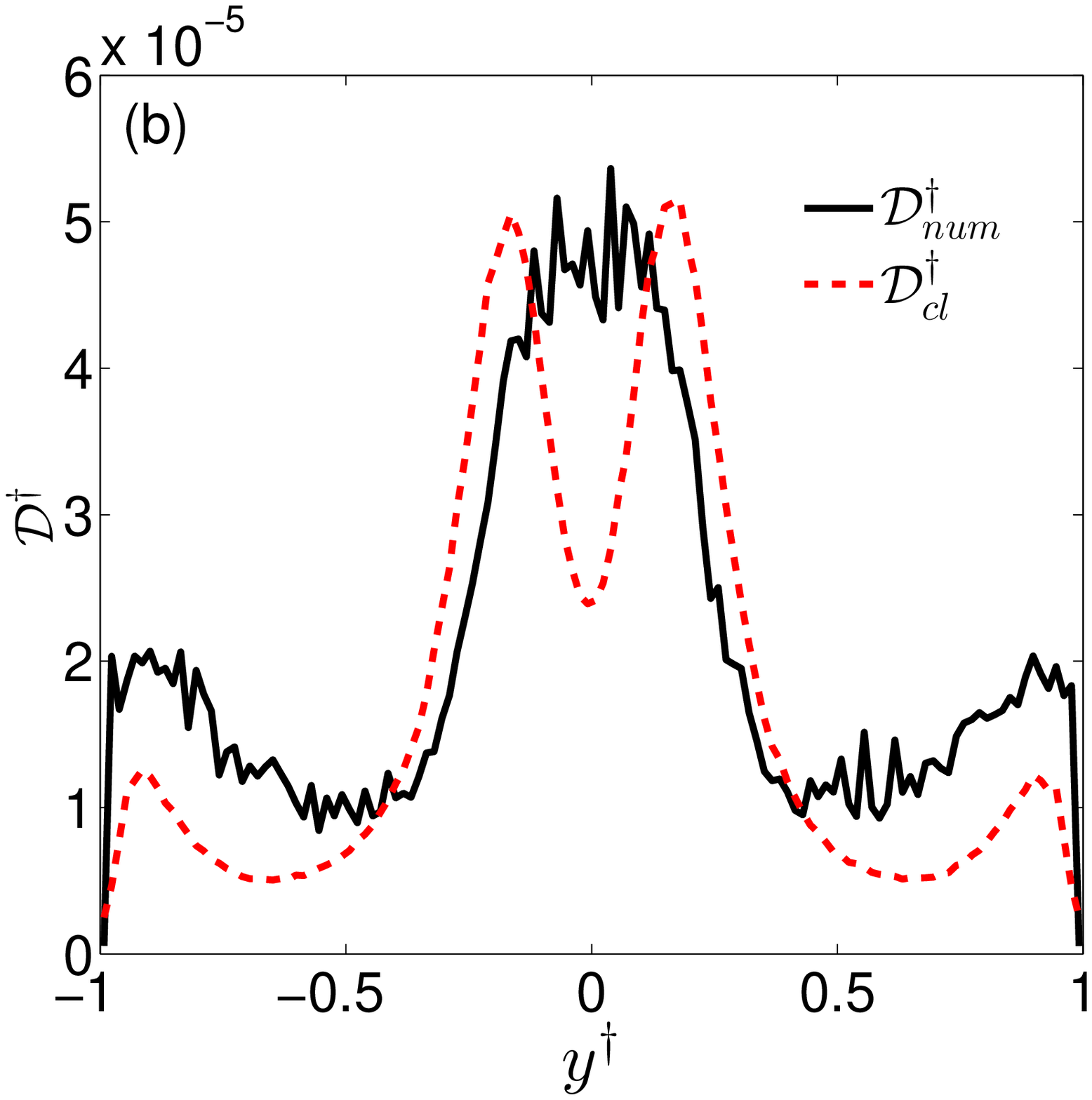} &   \includegraphics[scale=0.31 ]{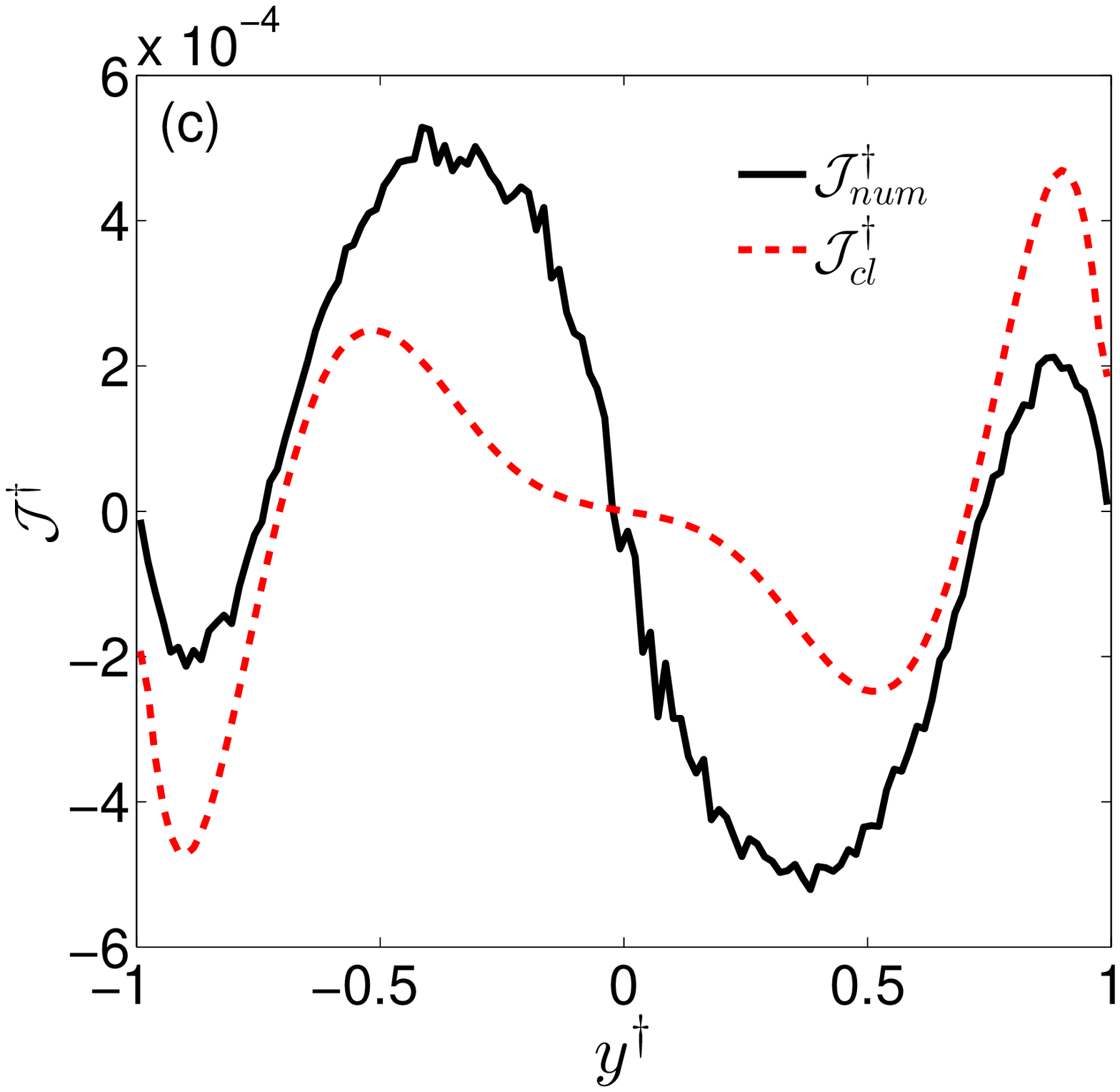}\\
     \end{tabular}
\caption{\label{f:3} For non-parabolic normal velocity profile: A comparison between production ($\C P$), decay ($\C D$) and flux ($\C J$) as obtained from numerical simulation and those obtained using their closure versions. Comparison of  the numerical data ( {\Large \bf -}\!{\Large \bf -}) to the various allowed production forms   of the Vinen equation (Panel a): ($-\cdot -$) --$\C P_1$, \Eq{eq:VinA}, (---) -- $\C P_2$, \Eq{eq:VinB} and ($-~-$) -- $\C P_3$, \Eq{P3B}.  Comparison   of  the numerical data for decay (Panel b) and flux term (panel c) with \Eq{eq:gen-balC} for $\C D$ and \Eq{flux} for $ \C J$ --  ($-~-$).}
\end{figure*}

\textbf{\emph{Numerical simulations}} were set up in a 3-dimensional planar channel geometry (see \Fig{f:1}a) of half-width $h$ with prescribed time-independent profile of the streamwise projection of the normal velocity $V_x\sp n(y)$.  To  find the vortex tangle configurations we used the vortex filament method, taking into account the potential flow $
 V_0\sp s $ to maintain the counterflow condition. Details of the   simulation method can be found in Refs.~\cite{b:1,b:Kond-1}. Here we used the reconnection method \cite{Samuels92} and the line resolution $\Delta \xi=1.6\times 10^{-3}$ cm.  Periodic conditions were used in the streamwise and spanwise  directions.  {Taking into account the fact that the boundary conditions for the superfluid component are still under discussion we adopt their simplest version: in the wall-normal $y$ direction    $V_y \sp s(\pm h)=0$ and ${\bm  s'(\pm h)}=(0,\pm 1,0)$ at the solid walls.}
  Periodically wrapped replicas of the tangle configuration were used in the  $x$- and $z$- directions, with reflected configurations in the $y$ direction.

 Having selected a stationary profile of $ V_x\sp n(y)\=V\sb n (y)$ we started with a set of arbitrary oriented circular vortex rings and solved the equation for the vortex line evolution. For the obtained dense vortex tangle  we found the time-averaged profile  $\<V_x\sp s(y)\>_t\= V\sb s(y)$.   In all cases we ran the simulations until we obtained steady mean profiles.  We take temperature $T=1.6$ K  with   mutual friction coefficients  $\alpha=0.098, \alpha'=0.016$,  Ref. \cite{DB98}.

\ We begin with a parabolic profile for $V\sb n(y)$. The  profiles of  $V\sb {ns}(y)$ and $\C L(y)$  in the dimensionless form:
 \begin{equation}\label{non-d}y^\dag\=  y/L\,, \ V^\dag\= V / \sqrt {\<  V\sb {ns}^2 \> }\,, \  \C L^\dag \= \kappa ^2 \C L/    \< V\sb{ns}^2 \>\,,
 \end{equation}
 are shown  in \Fig{f:1}b.
  Taking integrals in \Eqs{9} over the numerically found vortex tangle configuration we can compute the  production,  decay and flux terms, denoted as   $\C P\sb {num}(y,t)$,  $ \C D\sb {num}(y,t)$ and $\C J\sb {num}(y,t)$, respectively .  Then we compared them with the various their closure versions,  $\C P_1$, $\C P_2$, $\C P\sb {cl}$, $\C D\sb{cl}$ and $\C J\sb{cl}$.

 From the theoretical point of view the questions are: can we approximate  the integrals in \Eqs{9}   only in terms of the counterflow velocity $V\sb{ns}$ and $\C L$ itself, or would the integrals produce other dynamical variables that should require further coupled equations to close the system?  Is closure possible in general, or only in some conditions?

  Assuming that closure is allowed, dimensional considerations presented us with
different versions for the production term $\C P_n(y)$  given by \Eqs{eq:VinA}, (\ref{eq:VinB}) and (\ref{P3A}) for $n=1,2$ and 3.  With the widely accepted approximation\,\cite{b:1}  that the tangle curvature radius $R=1/ \widetilde S$
is proportional to the intervortex distance $\ell=1/\sqrt {\C L}$\,   one gets from \Eqs{dec} and\,\eqref{eq2A} the closure\,\eqref{eq:gen-balC} for   $\C D$ with
\begin{equation}\label{Cdec}
C\sb{dec}= c_2^2 \ln (R/a_0)/ 4\pi\,, \quad \mbox {where}\  c_2\equiv \ell \widetilde S\ .
 \end{equation}
 For the flux term\,\eqref{flux} we suggest  (in the  channel geometry):
 \begin{equation}\label{flux-c}
  \C J\sb {cl}(y,t)= -(\alpha / 2  \kappa) \, C\sb{flux}    \partial\, V\sb{ns}^2 /  \partial y \ .
 \end{equation}
In \Fig{f:2} we compare the numerical integrals\,\eqref{9} (shown as thick solid black lines) with  corresponding closures. The dot-dashed blue line in  \Fig{f:2}a shows the Vinen prediction, $\C P_1(y)$, \eqref{eq:VinA}, while the thin solid green line shows the alternative form $\C P_2(y)$, \Eq{eq:VinB}. By dashed  red line we show the prediction $\C P\sb{3}(y)$,which is evidently superior to the other two. From this data we can conclude that Eq. (\ref{P3B}) is the one that should be used in the present inhomogeneous case.

 Figure\,\ref{f:2}b shows that the numerical integral\,\eqref{prod} and the commonly used form\,\eqref{eq:gen-balC} for the decay term, $\C D$,  practically coincide,  meaning that $c_2$, defined by \Eq{Cdec},  is indeed $y$-independent. Figure~\ref{f:2}c also demonstrates very good agreement between   $\C J\sb{num}(y)$ and $\C J\sb{cl}(y)$ given by \Eqs{9a} and  \eqref{flux-c}.
 \begin{table}[h]
 \begin{tabular}{c c c c c  c c c  }
 \hline\hline
 &$V \sb n$   & $\sqrt{\< V\sb{ns}^2\>}/ V \sb n$ & $C_{\rm prod}$&   $C\sb{dec}$ &$C\sb{flux}$ &$C\sb{dec}\< V\sb{ns}^2\> $ &$C\sb{flux}/\< V\sb{ns}^2\> $  \\

 & cm/s &     &  &&&    (cm/s)$^2$   & (s/cm)$^2$

 \\ \hline

 1&1.0& 0.788   &  0.018 & 6.9    & 0.088 & 4.31   & 0.145 \\ 

 2&1.2& 0.785     & 0.018 & 4.9    & 0.124 & 4.35 & 0.147 \\ 

 3&1.5& 0.780  &0.019 & 3.2    & 0.210  & 4.32 & 0.146  \\   

 4&1&  1.247   & 0.022 &2.3    & 0.048  & 3.62 & 0.039  \\   
\hline\hline
 \end{tabular}
 \caption{\label{t:1} Closure constants for different values of the normal velocity maximum $V\sb n$   for parabolic   (lines 1-3) and non-parabolic (line 4) profiles. }

 \end{table}

 Realizing that the good match between the numerical data and \Eq{P3B} may be accidental due to particulary  chosen  numerical  parameters we repeated the simulations with other magnitudes of the counterflow velocities. We found again a good agreement between the numerical profiles $\C P\sb{num}(y)$, $\C P\sb{num}(y)$ and $\C J\sb{num}(y)$ with the corresponding closures, $\C P_3(y)$, $\C D\sb{cl}(y)$ and $\C J\sb{cl}(y)$. Consistency requires that the numerical constants $C\sb{prod}$, $C\sb{dec}$ and $C\sb{flux}$ in these closures should be $V\sb{ns}$ independent. Table\, \ref{t:1} shows that this is the case only for $C\sb {prod}$, while    $C\sb{dec}$ and $C\sb{flux}$ approximately depend on $V\sb{ns}$ as: $C\sb{dec}\propto 1/ \< V\sb{ns}^2\>$ and $C\sb{flux}\propto  \< V\sb{ns}^2\>$, where $\< V\sb{ns}^2\>$ is the mean-square of $V\sb{ns}$ across the channel (which, in its turn $\propto V\sb n^2$). From these facts we can conclude that our closure\,\eqref{P3A} for the production term is confirmed, while
the
 traditional closure\,\eqref{eq:gen-balC} for the decay term and simple closure\,\eqref{flux-c} seems to be questionable, although  they  reproduce well the profiles $\C D\sb{num}(y)$ and $\C J\sb{num}(y)$ in  the parabolic case.

 To clarify further the quality of the discussed closures we  imposed non-parabolic normal velocity profile  with two maxima and zero on the centerline, shown in \Fig{f:1}c. Although this profile looks strange, it may be realized in a counterflow experiment with non-homogeneous heating in short enough channel \cite{private:S}.
 For this profile we again computed the counterflow and vortex-line density profiles shown in \Fig{f:1}c. Next, we repeated all the steps described before, and found again   that our proposed form  Eq. (\ref{P3B}) fits the data  much better than the other two forms as seen in \Fig{f:3}a. Figures\,\ref{f:3}b,c show that  the closures for the decay and the flux terms, $\C D\sb{cl}$
 and $J\sb{cl}$, are running into trouble reflecting the numerical profiles only very roughly.  In particular this means that $c_2^2\equiv (\ell \widetilde S)^2$, which is involved in the closure\,\eqref{eq:gen-balC} via \Eq{Cdec}, varies across the channel by a factor of three, as follows from \Fig{f:3}b. Recall, that  in the parabolic case $c_2^2$ is $y$-independent, although  it depends on $\< V\sb{ns}^2\>$ approximately as  $1/ \< V\sb{ns}^2\>$. Consequently,  to improve the closure for the decay term  one needs to involve an additional balance equation for the  tangle curvature $\widetilde S$. Similarly, our analysis shows that to improve the closure for the    flux term,  one needs to involve information about the tangle anisotropy.

 \textbf{\emph{Conclusion}}:
 The suggested closure\,\eqref{P3A} for the production $\C P\sb{prod}(y)$ using  the counterflow velocity and the vortex line density profiles  can be considered as highly promising. However,  the closures  for the decay and flux are sensitive, and they may require accounting for additional tangle characteristics. The first candidate is the tangle curvature; the tangle anisotropy also can be important. Much more work in this direction is required to develop a consistent theory of the wall-bounded superfluid turbulence.

This paper had been  supported in
 part by Grant No. 14-29-00093 from Russian Science Foundation.
 L.K. acknowledges the kind hospitality at the Weizmann
 Institute of Science during the main part of the project.


\begin{thebibliography}{99}



\bibitem{b:a}  R.~J.  Donnelly, Quantized Vortices in Hellium II (Cambridge Univ. Press, Cambridge, UK, 1991).

\bibitem{b:b} Quantized Vortex Dynamics and Superfluid Turbulence, ed. by   C. F. Barenghi et al., Lecture Notes in Physics \textbf{571} (Springer-Verlag, Berlin, 2001).

\bibitem{b:Annett} A. F. Annett,  Superconductivity, Superfluids and Condensates (Oxford University Press, Oxford, 2004).

\bibitem{b:SS-2012}  L. Skrbek, K. R. Sreenivasan, Developed quantum turbulence and its decay,
 Phys. Fluids \textbf{24} 011301(2012).

\bibitem{nem} S. K. Nemirovskii, Quantum turbulence: Theoretical and numerical problems,
   Phys. Rep. {\bf 524,} 85 (2013).

\bibitem{b:PNAC2}  C. F. Barenghi,  V. S. L'vov and P.-E. Roche, Experimental, numerical, and analytical velocity spectra in turbulent quantum fluid,
Proc. Natl. Acad. Sci. USA {\bf 111,} (Supl.1), 4683 (2014).


\bibitem{b:PNAC1} C. F. Barenghi,  L. Skrbek  and K. R. Sreenivasan, Introduction to quantum turbulence,
 Proc. Natl. Acad. Sci. USA {\bf 111,} (Supl.1), 4647 (2014).

 \bibitem{Helsinki-rev} V.B. Eltsov, R. de Graaf, R. Hanninen, M. Krusius, R.E. Solntsev, V.S. L'vov, A.I. Golov, P.M. Walmsley, Turbulent dynamics in rotating helium superfluids, Progress in Low Temperature Physics, XVI pp. 46-146 (2009).

\bibitem{b:PNAC3}  E. Fonda,  D. P. Meichle,   N. T. Ouellette, S. Hormoz,   and D. P. Lathrop, Direct observation of Kelvin waves excited by quantized vortex reconnection, Proc. Natl. Acad. Sci. USA {\bf 111,} (Supl.1), 4707 (2014).

\bibitem{b:PNAC4} W. Guo,  M. La Mantiac, D. P. Lathrop   and S.W. Van Sciver,  Visualization of two-fluid flows of superfluid helium-4,
Proc. Natl. Acad. Sci. USA {\bf 111,} (Supl.1), 4653 (2014).

\bibitem{b:11} L. Skrbek, A.V. Gordeev, F. Soukup, Decay of counterflow He II turbulence in a finite channel: Possibility of missing links between classical and quantum turbulence, Phys. Rev. E \textbf{67}, 047302 (2003).

\bibitem{B:12} S.W. Van Sciver, S. Fuzier, and T. Xu, J., Particle Image Velocimetry Studies of Counterflow Heat Transport in Superfluid Helium II,
J. Low Temp. Phys. \textbf{148}, 225 (2007).

\bibitem{B:13} T. Zhang and S. W. Van Sciver,  Large-scale turbulent flow around a cylinder in counterflow superfluid 4He  (He (II)),
Nat. Phys. \textbf{1}, 36 (2005).

\bibitem{B:14} G. P. Bewley, D. P. Lathrop, and K. R. Sreenivasan, Superfluid Helium: Visualization of quantized vortices, Nature (London) \textbf{441}, 588 (2006).

\bibitem{B:15} M. S. Paoletti, R. B. Fiorito, K. R. Sreenivasan, and D. P. Lathrop,Visualization of Superfluid Helium Flow, J. Phys. Soc. Jap. \textbf{77}, 111007 (2008).

\bibitem{B:19} D. N. McKinsey, W. H. Lippincott, J. A. Nikkel, and W. G. Rellergert, Trace Detection of Metastable Helium Molecules in Superfluid Helium by Laser-Induced Fluorescence, Phys. Rev. Lett. 95, 111101 (2005).

\bibitem{B:20} W. G. Rellergert, S. B. Cahn, A. Garvan, J. C. Hanson, W. H. Lippincott, J. A. Nikkel, and D. N. McKinsey, Detection and Imaging of He2 Molecules in Superfluid Helium, Phys. Rev. Lett. \textbf{100}, 025301 (2008).

\bibitem{B:21} W. Guo, J. D. Wright, S. B. Cahn, J. A. Nikkel, and D. N. McKinsey, Metastable Helium Molecules as Tracers in Superfluid He4, Phys. Rev. Lett. \textbf{102}, 235301 (2009).

\bibitem{B:22} W. Guo, S. B. Cahn, J. A. Nikkel, W. F. Vinen, and D. N. McKinsey, Visualization Study of Counterflow in Superfluid He4 using Metastable Helium Molecules
 Phys. Rev. Lett. \textbf{105}, 045301 (2010).

 \bibitem{B:23}  A. Marakov, J. Gao, W. Guo, S. W. Van Sciver, G. G. Ihas, D. N. McKinsey, and W. F. Vinen,
     Visualization of the normal-fluid turbulence in counterflowing superfluid He4.
Phys. Rev. B \textbf{91}, 094503 (2015).

\bibitem{b:17} C.F. Barenghi, A.V. Gordeev, L. Skrbek, Depolarization of decaying counterflow turbulence in He II, Phys. Rev. E \textbf{74}, 026309 (2006).

\bibitem{b:18} M. Sciacca, Y.A. Sergeev, C.F. Barenghi, L. Skrbek, Saturation of decaying counterflow turbulence in helium II, Phys. Rev. B \textbf{82}, 134531 (2010).

\bibitem{b:PNAC5}   N. G. Berloff,  M. Brachet and  N. P. Proukakis,  Modeling quantum fluid dynamics at nonzero temperatures,
 Proc. Natl. Acad. Sci. USA {\bf 111,} (Supl.1), 4675 (2014).

\bibitem{b:PNAC6} R. H\"{a}nninen  and   A. W. Baggaley, Vortex filament method as a tool for computational visualization of quantum turbulence,
 Proc. Natl Acad. Sci. USA {\bf 111,} (Supl.1), 4667 (2014).

 \bibitem{B:9} A. W. Baggaley and Laizet,  Vortex line density in counterflowing He II with laminar and turbulent normal fluid velocity profiles,
Phys. Fluids \textbf{25}, 115101 (2013).
\bibitem{b:Kond-1} L. Kondaurova,   V. L'vov,  A. Pomyalov  and I. Procaccia, Structure of a quantum vortex tangle in $^4$He counterflow turbulence,
Phys. Rev. B   \textbf{89,} 014502 (2014).

\bibitem{b:PNAC7} G. V. Kolmakov,  P. V. E. McClintock and S. V. Nazarenko,  Wave turbulence in quantum fluids, Proc. Natl. Acad. Sci. USA {\bf 111,} (Supl.1), 4727 (2014).

\bibitem{b:A-10}  H. Adachi, S. Fujiyama, M. Tsubota, Steady-state counterflow quantum turbulence: Simulation of vortex filaments using the full Biot-Savart law,
Phys. Rev. B \textbf{81}, 104511 (2010).

\bibitem{LNV-2004} V. S. L'vov, S. V. Nazarenko and G. E. Volovik, Energy spectra of developed superfluid turbulence, J. Low Temp. Phys. \textbf{80}, 535  (2004).

\bibitem{LNS-2006} V.S. L'vov, S.V. Nazarenko and L. Skrbek, Energy Spectra of Developed Turbulence in Helium Superfluids, J. Low Temp. Phys. \textbf{145}, 125 (2006).

\bibitem{LNR-2008}  V. S. L'vov, S. V. Nazarenko and O. Rudenko, Gradual eddy-wave crossover in Superfluid turbulence, J. of Low Temp. Phys. 153, 140-161 (2008).

\bibitem{BLPP-2012} L. Boue, V.S. L'vov, A. Pomyalov, I. Procaccia, Energy spectra of superfluid turbulence in He-3B,  Phys. Rev. B \textbf{85}, 104502 (2012).

\bibitem{BLPP-2013} L. Boue, V.S. L'vov, A. Pomyalov, I. Procaccia,  Enhancement of intermittency in superfluid turbulence,  Phys. Rev. Lett.,\textbf{ 110}, 014502 (2013).

\bibitem{BLNNPP-2015} L. Boue, V S. L'vov, Y. Nagar, S. V. Nazarenko, A. Pomyalov, and I. Procaccia, Energy and Vorticity Spectra in Turbulent Superfluid 4He from T = 0 to T$_\lambda$, Phys. Rev. B \textbf{91}, 144501 (2015).

\bibitem{b:L-41}  L.D. Landau, Theory of superfluidity of helium-II, J. Phys. USSR  \textbf{5}, 71 (1941).

\bibitem{b:T-40}  L. Tisza, J. Phys. Radium  \textbf{1}, 164 (1940) , \emph{ibid} \textbf{1,} 350 (1940).

\bibitem{HV} H.E. Hall, W.F. Vinen, The rotation of liquid helium II. II. The theory of mutual friction in uniformly rotating helium II,
 Proc. R. Soc. Lond. A Math Phys. Sci. \textbf{238}, 215 (1956).

\bibitem{BK} I.L. Bekarevich, I.M. Khalatnikov,
 Phenomenological Derivation of the Equations of Vortex Motion in He II,
 Zh. Eksp. Teor. Fiz. \textbf{40}, 920 (1961) (Sov. Phys. JETP \textbf{13}, 643 (1961)).

\bibitem{b:1}  K.W. Schwarz, Three-dimensional vortex dynamics in superfluid $^4$He: Homogeneous superfluid turbulence,
 Phys. Rev. B  \textbf{38,}  2398 (1988).


\bibitem{b:5}  W. F. Vinen, Mutual friction in a heat current in liquid helium II. I. Experiments on steady heat currents,
Proc. R. Soc. Lond. A Math. Phys. Sci. \textbf{ 240,} 114 (1957).


\bibitem{b:Kop} N. B. Kopnin.
 Vortex Instability and the Onset of Superfluid Turbulence
Phys. Rev. Lett. \textbf{92}, 135301 (2004).

 \bibitem{b:6} W. F. Vinen, Mutual friction in a heat current in liquid helium II. II. Experiments on transient,
Proc. R. Soc. Lond. A Math. Phys. Sci. \textbf{240,} 128 (1957).


\bibitem{b:8}  S. K. Nemirovskii  and W. Fiszdon, Chaotic quantized vortices and hydrodynamic processes in superfluid helium,
 Rev. Mod. Phys.  \textbf{ 67,}  37 (1995).

\bibitem{BB} Y.M. Bunkov, A. I. Golov, V. S. L'vov, A. Pomyalov and I. Procaccia,  Evolution of Neutron-Initiated Micro-Big-Bang in superfluid He 3B, Phys. Rev B \textbf{90}, 024508 (2014).

\bibitem{Samuels92}  D. C. Samuels ,  Velocity matching and Poiseuille pipe flow of superfluid helium,
Phys. Rev. B {\bf 46,} 11714 (1992).

\bibitem{DB98}  R. J. Donnelly  , C. F. Barenghi , The Observed Properties of Liquid Helium at the Saturated Vapor Pressure,
 J. Phys. Chem. Ref. Data \textbf{27,} 1217(1998).

\bibitem{private:S} Skrbek, L., \textit{private communication.}




\end{thebibliography}
 \end{document}